\documentclass[11pt]{article}
\usepackage{latexsym}
\usepackage{amsmath}
\usepackage{amssymb}
\usepackage{bbm}
\usepackage{graphicx}
\usepackage{subfigure}
\usepackage{dcolumn}
\usepackage{bm}
\usepackage{cite}
\usepackage{color}
\usepackage[colorlinks,linkcolor=magenta,anchorcolor=blue,citecolor=blue]{hyperref}
\usepackage{tikz}
\usepackage{geometry}
\usetikzlibrary{arrows} 
\usepackage{fancyhdr}
\usepackage{xcolor}
\newsavebox\MBox

\title{Supersymmetric Localization in GLSMs for Supermanifolds}
\author{Wei Gu\footnote{weig8@vt.edu}, Hao Zou\footnote{hzou@vt.edu} \\ Department of Physics, Virginia Tech \\ 850 West Campus Dr., Blacksburg, VA 24061}
\date{July 2018}
\begin{document}
\maketitle
\begin{abstract}
	In this paper we apply supersymmetric localization to study 
gauged linear sigma models (GLSMs) describing supermanifold target spaces. We use the localization method to show that A-twisted GLSM correlation functions for certain supermanifolds are equivalent to A-twisted GLSM correlation functions for hypersurfaces in ordinary spaces under certain conditions. We also argue that physical two-sphere partition functions are the same for these two types of target spaces.
Therefore, we reproduce the claim of \cite{Sethi:1994ch,Schwarz:1995ak}. Furthermore, 
we explore elliptic genera and (0,2) deformations and find similar phenomena.
\end{abstract}

\newpage
\tableofcontents

\newpage
\section{Introduction}
\label{sec:introduction}

Supermanifolds have recently been of interest in the community, see {\it e.g.} \cite{Donagi:2014hza,Donagi:2013dua,Witten:2013cia,Witten:2012bh,Witten:2012ga,Witten:2012bg,Jia:2016jlo,Jia:2016rdn,Seki:2005hx,Seki:2006cj}. The purpose of this paper is to use supersymmetric localization to explore properties of gauged linear sigma models with target supermanifolds, checking the equivalence \cite{Sethi:1994ch,Schwarz:1995ak} of 
A-twisted nonlinear sigma models (NLSMs) on supermanifolds with A-twisted nonlinear sigma models on ordinary hypersurfaces and complete
intersections in (2,2) supersymmetric cases. We check this claim by directly comparing A-twisted correlation functions for both sides. Furthermore, 
we also compare elliptic genera as well as partition functions on two-sphere. We also discuss analogues for (0,2) supersymmetric theories.

In this paper, we require that NLSMs on toric supermanifolds have non-negative beta functions, which is equivalently to require these supermanifolds have non-negative super-first Chern Classes. (For example, the super-first Chern class of $\mathbb{C}\mathbb{P}^{N|M}$ can be non-negative when $N+1 \geq M$ in Eq. (3.6) of \cite{Sethi:1994ch}.) Therefore we could use GLSMs as the UV-complete theories to study these supermanifolds. GLSMs for supermanifold target spaces have been studied in \cite{Seki:2005hx,Seki:2006cj}. In this paper, we apply supersymmetric localization to study GLSMs. This is a powerful tool for ordinary GLSMs \cite{Closset:2015rna,Benini:2015noa,Closset:2015ohf,Benini:2016qnm}, which we extend to GLSMs for supermanifolds. The philosophy of supersymmetric localization is to do calculations at worldsheet UV for some RG-invariant quantities, such as correlation functions of the topological field theory. 

Using supersymmetric localization, we calculate the correlation functions of A-twisted GLSMs for supermanifolds and we show that they match with A-twisted GLSMs for certain ordinary hypersurfaces and complete intersections. In addition, we also find that physical two-sphere partition functions for supermanifolds and for corresponding ordinary manifolds are equal. Therefore, we conjecture that the mirror maps are the same for both sides \cite{Jockers:2012dk,Gomis:2012wy}. However, one subtlety is that some properties of supermanifolds are not quite clear. We leave the proof of this conjecture as future work.

In section \ref{sec:glsm}, we briefly review GLSMs for ordinary toric varieties via concrete examples with three different target spaces: $\mathbb{C}\mathbb{P}^4$, $\rm{Tot}\left(\mathcal{O}(-d)\rightarrow\mathbb{C}\mathbb{P}^4\right)$ and a hypersurface of degree $d$ in $\mathbb{C}\mathbb{P}^4$. We focus on the correlation function calculations via supersymmetic localization, which will be used in section~\ref{sec:comparison}. 

In section \ref{sec:glsmforsupermanifolds}, we discuss GLSMs for supermanifolds. The general description is 
based on \cite{Seki:2005hx,Seki:2006cj}, but we do not consider superpotentials for supermanifolds in this paper 
because it is not relevant for reproducing the claims in \cite{Sethi:1994ch,Schwarz:1995ak}, which in principle are the focus
of this paper.  In section \ref{sec:chrialring}, the general chiral ring relations of supermanifolds will be obtained following \cite{Morrison:1994fr,Hori:2003ic}. Then using supersymmetric localization \cite{Closset:2015rna,Benini:2015noa,Benini:2013nda,Benini:2013xpa}, formulas for correlation functions and elliptic genera are also given.

In section \ref{sec:comparison}, we apply those formulas given in last section to several examples and the statement in \cite{Sethi:1994ch,Schwarz:1995ak} can be obtained immediately under certain conditions. This statement is for Higgs branches, but our calculations are all done on Coulomb branches, for example the correlation functions, (\ref{eq:local}) and (\ref{eq:crs}). However, the correlation functions on the Higgs branch are equivalent to the correlation functions on the Coulomb branch when turn off twisted masses. 

In section \ref{sec:partitionfunction}, we study the two-sphere 
partition functions of physical $(2,2)$ theories. 
We find that the partition functions for 
certain supermanifolds are equivalent to the partition functions 
for hypersurfaces in ordinary spaces. 
In section \ref{sec:(0,2)}, we study $(0,2)$ deformation of 
$(2,2)$ theories. We generalize the story in \cite{Closset:2015ohf} 
to GLSMs for supermanifolds without $(0,2)$ superpotentials. 
Since there are no $J$-terms,
there should no constraints for $E$-deformations. 
However, the $E$-deformations for $(0,2)$ GLSMs for hypersurfaces have 
certain constraints due to supersymmetry, see {\it e.g.} 
Eq.~(\ref{eq:susyconstraint1}) and Eq.~(\ref{eq:susyconstraint2}). 
Therefore, the $(0,2)$ version of that statement \cite{Sethi:1994ch,Schwarz:1995ak} only hold 
for deformations obeying certain constraints.

\section{Review of GLSMs for Toric Varieties}
\label{sec:glsm}

In this section, we briefly review some aspects of GLSMs for toric varieties 
and how to compute correlation functions via supersymmetric localization on 
the Coulomb branch in some concrete examples. 

Consider a GLSM with gauge group $U(1)^k$ and $N$ chiral superfields $\Phi_i$ 
of gauge charges $Q_i^a$ and vector R charges\footnote{
In order to make this GLSM to be A-twistable on a two-sphere, 
the vector R-charges, denoted as $R_V$, should be integers 
\cite{Guffin:2008kt,Hori:2013ika}.} 
$R_i$, 
where $a = 1, \dots k$ and $i = 1, \ldots, N$. The lowest components of $\Phi_i$ are bosonic scalars $\phi_i$, 
and we call these $\Phi_i$ {\it even} chiral superfields. 
The Lagrangian and general discussions of this model can be found 
in the literature, see {\it e.g.} \cite{Witten:1993yc,Morrison:1994fr}. 

In the geometrical phase of this GLSM, the vacuum moduli space could have the flavor symmetry of the form
\begin{equation}
\label{eq:globalsymm}
	\times_{\alpha}U(N_\alpha)/U(1)^k,
\end{equation}
where each $N_\alpha$ is the number of chiral superfields which have the same gauge charge and $\sum_{\alpha}N_\alpha = N$. However, if the theory has a superpotential, the flavor symmetry will be smaller \cite{Nekrasov:2009uh}. For example, in the GLSM for the quintic, the vacuum moduli space has no continuous isometry. \footnote{For the GLSM for a hypersurface, one can still do computations on the Coulomb branch by lifting the chiral superfields \cite{Hori:2011pd}.}

The correlation function for a general operator $\mathcal{O}(\sigma)$ can be 
calculated via localization on the Coulomb branch as \cite{Closset:2015rna}
\begin{equation}
\label{eq:local}
	\left<\mathcal{O}(\sigma)\right> = (-1)^{N_*} \sum_{\bm m}\oint_{\rm JK-Res} \prod_{a=1}^k \left( \frac{{\rm d}\sigma_a}{2\pi i}\right) \mathcal{O}(\sigma) Z_{\bm m}^{\rm 1-loop} q^{\bm m},
\end{equation}
where $q^{\bm m}= e^{-t^a m_a}$, in which:
\begin{align*}
	t^a &= r^a - i\theta^a,	\\
	r^a &=  r_0^a + \sum_i Q_i^a \ln \frac{\mu}{\Lambda},
\end{align*}
and $Z_{\bm m}^{\rm 1-loop}$ is the one loop determinant. 
For abelian gauge theories, it is known that
\begin{equation*}
	Z_{\bm m}^{\rm 1-loop} = \prod_i (Q_i^a \sigma_a + \tilde{m}_i)^{R_i-1-Q_i(\bm m)},
\end{equation*}
in which
\begin{equation*}
	Q_i({\bm m}) = Q_i^a m_a,
\end{equation*}
and $\tilde{m}_i$ are the twisted masses associated to the flavor symmetry. The overall factor $(-1)^{N_*}$, where $N_*$ is the number of $p$ fields, comes from the assignment for the fields with R-charge $2$ \cite{Closset:2015rna,Morrison:1994fr}. We will later see this overall factor would automatically show up from the redefinition of $q$'s in the supermanifold case in following sections. The special case, $N_* = 0$, corresponds to target space without a superpotential.

Next, we will apply the above formula to calculate several concrete examples.
\vskip 5mm
\noindent \underline{GLSM for $\mathbb{C}\mathbb{P}^4$}

In this model, we have five chiral superfields with $U(1)$ charges and $R_V$-charges given by
\begin{equation*}
\arraycolsep=4.0pt \def\arraystretch{1.2}
\begin{array}{|c|ccccc|}
\hline
	Q & 1 & 1 & 1 & 1 & 1\\ \hline
	R_V & 0 & 0 & 0 & 0 & 0 \\ \hline
\end{array}
\end{equation*}
and it has a $SU(5)$ flavor symmetry. 
For simplicity, we set twisted masses to zero.

Then from the formula (\ref{eq:local}), we obtain:
\begin{equation*}
	\left<\mathcal{O}(\sigma)\right> = \sum_{k \geq 0} \oint \frac{{\rm d}\sigma}{2\pi i} \frac{\mathcal{O}(\sigma)}{\sigma^{5+5k}} q^k.
\end{equation*}
If take $\mathcal{O}(\sigma) = \sigma^{5k+4}$, we could immediately obtain
\begin{equation*}
	\left<\sigma^{5k+4}\right> = q^k,
\end{equation*}
and this equation encodes the chiral ring relation as
\begin{equation*}
	\sigma^5 = q.
\end{equation*}

\vskip 5mm
\noindent \underline{GSLM for $\rm{Tot}\left(\mathcal{O}(-d)\rightarrow\mathbb{C}\mathbb{P}^4\right)$}

$\rm{Tot}\left(\mathcal{O}(-d)\rightarrow\mathbb{C}\mathbb{P}^4\right)$ is the 
total space of the bundle $\mathcal{O}(-d)\rightarrow\mathbb{C}\mathbb{P}^4$. 
For the special case when $d=5$, it is also called $V^+$ model as in \cite{Morrison:1994fr}. In this example, we have six chiral superfields with $U(1)$ charges and $R_V$-charges given by
\begin{equation*}
\arraycolsep=4.0pt \def\arraystretch{1.2}
\begin{array}{|c|cccccc|}
\hline
	Q & 1 & 1 & 1 & 1 & 1 & -d\\ \hline
	R_V & 0 & 0 & 0 & 0 & 0 & 0 \\ \hline
\end{array}
\end{equation*}
This model has the flavor symmetry $SU(5) \times U(1)$. We require $\sum_i Q_i \geq 0$ so this system has a geometric phase corresponding to a weak coupling limit. Then we have
\begin{align*}
	\left< \mathcal{O}(\sigma) \right> &= \sum_{k} \oint_{JK-Res} \frac{{\rm d}\sigma}{2\pi i} \frac{\mathcal{O}(\sigma)}{\sigma^{5+5k}(-d\sigma)^{1-dk}} q^k,\\
	&= \sum_{k \geq 0} \oint \frac{{\rm d}\sigma}{2\pi i} \frac{\mathcal{O}(\sigma)}{\sigma^{6+(5-d)k} (-d)^{1-dk}}q^k.
\end{align*}
For the special case $d = 5$, we can further obtain the following chiral ring relation: 
\begin{equation*}
	\left< \sigma^5 \right>  = -\frac{1}{5} \frac{1}{1+5^5 q}.
\end{equation*}

\vskip 5mm
\noindent \underline{GLSM for the Hypersurface in $\mathbb{C}\mathbb{P}^4$}

This model is defined by six chiral superfields with $U(1)$ charges and $R_V$-charges given by:
\begin{equation*}
\arraycolsep=4.0pt \def\arraystretch{1.2}
	\begin{array}{|c|cccccc|}
	\hline
		Q & 1 & 1 & 1 & 1 & 1 & -d \\ \hline
		R_V & 0 & 0 & 0 & 0 & 0 & 2 \\ \hline
	\end{array}
\end{equation*}
which we also require $\sum_i Q_i \geq 0$. It has no flavor symmetry.  We have
\begin{align*}
	\left< \mathcal{O}(\sigma)  \right> &= (-1)^1\sum_{k} \oint_{JK-Res} \frac{{\rm d}\sigma}{2\pi i} \frac{\mathcal{O}(\sigma)(-d\sigma)^2}{\sigma^{5+5k}(-d\sigma)^{1-dk}} q^k,\\
	&= -\sum_{k \geq 0}\oint \frac{{\rm d}\sigma}{2\pi i} \frac{\mathcal{O}(\sigma)(-d )^{1+dk}}{\sigma^{4+(5-d)k}} q^k.
\end{align*}
In particular, if $d=5$, then it satisfies the Calabi-Yau condition. Then,
\begin{align*}
	\left< \mathcal{O}(\sigma)  \right> = -\sum_{k\geq 0 }\oint \frac{{\rm d}\sigma}{2\pi i} \frac{\mathcal{O}(\sigma)(-5 )^{1+5k}}{\sigma^{4}} q^k.
\end{align*}
Take $\mathcal{O}(\sigma) = \sigma^{3}$, then we can obtain
\begin{equation*}
	\left< \sigma^{3}  \right> = \frac{5}{1+5^5q}.
\end{equation*}
This correlation function is in agreement with $\left< \sigma^{3}\left(-(-5\sigma)^2\right) \right>$ in the previous $V^+$ model \cite{Morrison:1994fr}.

\section{GLSMs for Complex K\"{a}hler Supermanifolds}
\label{sec:glsmforsupermanifolds}

A supermanifold $X$ of dimension $N|M$ is locally described by $N$ even coordinates and $M$ odd coordinates together with compatible transition functions. If it is further a split supermanifold, then it can be viewed as the total space of an odd vector bundle $V$ of rank $M$ over a $N$-dimensional manifold, which is along the even directions and denoted $X_{\rm red}$: 
$$X \simeq {\rm Tot}(V \rightarrow X_{\rm red}).$$
For more rigorous definitions of supermanifolds and split supermanifolds, 
we recommend \cite{Witten:2012bg}. According to the fundamental structure theorem\cite{Witten:2012bg}, every smooth supermanifold can be split, so even the split case is still considerable. 

To build up a $(2,2)$ GLSM as a UV-complete theory of a NLSM for a 
complex K\"{a}hler supermanifold $\mathcal{M}$, 
we only consider those toric supermanifolds \cite{Schwarz:1995ak} 
obeying certain constraints, which we will give later as 
Eq.~(\ref{eq:nlsmcd}). We obtain this from the GLSM perspective, but it can be derived from NLSMs \cite{Sethi:1994ch}. By toric supermanifold, we mean that $\mathcal{M}$ has a complexified symmetry group $(\mathbb{C}^*)^k$ and can be obtained as a symplectic reduction of a super vector space by an abelian gauge group, which is realized in a GLSM by gauging a group action on a super vector space (corresponding to matter fields). It was pointed out in \cite{Schwarz:1995ak} that this kind of supermanifold is also split. 
Therefore, we can still take advantage of the bundle structure of 
split supermanifolds in our construction. One example of these toric supermanifolds is $\mathbb{C}\mathbb{P}^{4|1}$, which is defined by
\begin{equation}
\label{eq:cp41}
	\left\{ [x_1,x_2,x_3,x_4,x_5, \theta] \ |\ (x_1,x_2,x_3,x_4,x_5, \theta) \sim (\lambda x_1, \lambda x_2, \lambda x_3, \lambda x_4, \lambda x_5, \lambda^d \theta) \right\}.
\end{equation}
This is a different geometry than
$\mathbb{C}\mathbb{P}^4$.  For example, 
on $\mathbb{C}\mathbb{P}^{4|1}$ we can choose a patch where 
$\{ x_1, \ldots, x_5\}$ all vanish, while the odd coordinate is nonzero.

\subsection{The Model}
\label{sec:themodel}

In order to construct the GLSM for a toric supermanifold 
described by a $U(1)^k$ gauge theory, 
we can follow the construction of $V^{+}$ model \cite{Morrison:1994fr}  but change the statistical properties along the bundle directions. 
In other words, we view fields along bundle direction as ghosts. 
In \cite{Seki:2005hx}, there is a formal discussion about building GLSMs for supermanifolds. 
Here we only focus on toric supermanifolds. More specifically, we have two sets of chiral superfields:
\begin{itemize}
	\item $N+1$ {\it (Grassmann) even chiral superfields} $\Phi_i$ with $U(1)^k$ gauge charges $Q_i^a$ and R-charges $R_i$, whose lowest components are bosonic  scalars;
	\item $M$ {\it (Grassmann) odd chiral superfields} $\tilde{\Phi}_{\mu}$ with gauge $U(1)^k$ charges $\tilde{Q}_\mu^a$  and R-charges $\tilde{R}_{\mu}$, whose lowest components are fermionic scalars. \footnote{For general discussions, we use tilde `$\sim$' to indicate the odd chiral superfields and their charges.}
\end{itemize}
In the above, we impose an analogue of a Fano requirement for the
supermanifold, requiring that for each index $a$ 
\begin{equation}
\label{eq:nlsmcd}
\sum_{i}Q_i^a - \sum_{\mu}\tilde{Q}_{\mu}^a \geq 0,
\end{equation}
and in later sections we impose this condition implicitly.
(We will derive this condition from the worldsheet beta function later
in this section.)

Associated to the gauge group $U(1)^k$, 
there are $k$ vector superfields: $V_a,\ a =1,\ldots,k $. 
The total Lagrangian consists of five parts\footnote{
For a comprehensive expression for the Lagrangian, 
please refer to \cite{Seki:2005hx}[section 2]. }:
\begin{equation*}
	\mathcal{L} = \mathcal{L}_{\rm kin}^{\rm even} + \mathcal{L}_{\rm kin}^{\rm odd} + \mathcal{L}_{\rm gauge}+ \mathcal{L}_{W}+\mathcal{L}_{\tilde{W}} .
\end{equation*}
As advertised in the introduction, we will consider a vanishing superpotential in this paper, i.e. $W = 0$. Take the classical twisted superpotential to be a linear function\footnote{
We use notations of \cite{Hori:2003ic}.}
\begin{equation}
\label{eq:twistpotential}
	\tilde{W} = -\sum_a t^a\Sigma_a.
\end{equation}
In the above Lagrangian, the even kinetic part, the gauge part and the 
twisted superpotential part share the same form as in a GLSM 
for an ordinary target space. The odd kinetic part is defined in the same fashion as the even part \cite{Seki:2005hx}:
\begin{equation}
\label{eq:oddkin}
	\mathcal{L}_{\rm kin}^{\rm odd} = \int d^4\theta \sum_{\mu}\bar{\tilde{\Phi}}_{\mu}e^{2Q_\mu^a V_a}\tilde{\Phi}_{\mu}.
\end{equation}
The equations of motion for the auxiliary fields $D^a$ inside vector superfields are 
\begin{equation}
\label{eq:dterm}
	D^a = -e^2 \left( \sum_i Q_i^a |\phi_i|^2 + \sum_{\mu} \tilde{Q}_{\mu}^a 
 \overline{\tilde{\phi}}_{\mu} \tilde{\phi}_{\mu} - r^a\right),
\end{equation}
where $r^a$ are the FI parameters. Since $W=0$, the equations of motion for the auxiliary fields $F_{i/\mu}$ inside even/odd chiral superfields are
\begin{equation*}
	F_i = 0,\quad F_{\mu} = 0.
\end{equation*}
The potential energy is
\begin{equation*}
	U = \frac{1}{2e^2}D^2 + |\sigma|^2 \left( \sum_i Q_i^2 |\phi_i|^2 + 
\sum_{\mu}\tilde{Q}_{\mu}^2 \overline{\tilde{\phi}}_{\mu}
\tilde{\phi}_{\mu} \right).
\end{equation*}

Semiclassically, we can discuss low energy physics by requiring $U=0$, 
{\it i.e.} $\sigma = 0$ and $D = 0$, which is 
\begin{equation*}
	 \sum_i Q_i^a |\phi_i|^2 + \sum_{\mu} \tilde{Q}_{\mu}^a 
\overline{\tilde{\phi}}_{\mu} \tilde{\phi}_{\mu} - r^a = 0.
\end{equation*}
In the case with one $U(1)$, we often require a geometric phase where $r\gg 0$ defined by $(\mathbb{C}^{N+1|M} - Z) / \mathbb{C}^*$.\footnote{To be thorough, we also need to define theory at other phases. For example, there exists another phase called nongeometric phase corresponding to $r \leq 0$ \cite{Aganagic:2004yh}. However, supersymmetric localization are calculated at the worldsheet UV, which corresponds to a geometric phase in this paper under the condition Eq. (\ref{eq:nlsmcd}).  }  Returning to the general case, in the phase $r^a \gg 0$ for all $a\in \{1, \dots, k \}$, 
the above condition requires that not all $\phi_i$ or $\tilde{\phi}_\mu$  can vanish, 
then the target space is a super-version of the toric variety, $X$, 
which we call a super toric variety:
\begin{equation}
\label{eq:supertoricvariety}
	X \simeq \frac{\mathbb{C}^{N+1|M}-Z}{\left(\mathbb{C}^*\right)^k},
\end{equation}
where the torus action $\left(\mathbb{C}^*\right)^k$ is defined as, for each $a$,
\begin{equation*}
	\left(\dots ,\phi_i,\dots, \tilde{\phi}_{\mu},\dots  \right) \mapsto \left(\dots ,\lambda^{Q_i^a}\phi_i, \dots, \lambda^{\tilde{Q}_{\mu}^a}\tilde{\phi}_{\mu},\dots  \right), \quad \lambda\in\mathbb{C}^*.
\end{equation*}

As in the case for ordinary toric varieties, we have symmetries for super toric varieties. For a general case, (\ref{eq:supertoricvariety}), the maximal torus of the symmetry would be:
\begin{equation*}
	\frac{U(1)^{N+1}\times U(1)^M}{U(1)^k}.
\end{equation*}
Since we are not considering superpotentials in our models, 
this symmetry will not break.

The one-loop correction to the $D$-terms can be calculated as in \cite{Seki:2005hx}:
\begin{equation}
	\left< - \frac{D^a}{e^2} \right>_{\rm 1-loop} = \frac{1}{2} \sum_i Q_i^a \ln\left( \frac{\Lambda^2}{Q_i^b Q_i^c\bar{\sigma}_b\sigma_c} \right) - \frac{1}{2} \sum_{\mu} \tilde{Q}_{\mu}^a \ln\left( \frac{\Lambda^2}{\tilde{Q}_{\mu}^b\tilde{Q}_{\mu}^c\bar{\sigma}_b\sigma_c} \right).
\end{equation}
Therefore, the effective FI-parameters are given as
\begin{align*}
	r^a_{\rm eff} &= r^a -\frac{1}{2} \sum_i Q_i^a \ln\left( \frac{\Lambda^2}{Q_i^b Q_i^c\bar{\sigma}_b\sigma_c} \right) - \frac{1}{2} \sum_{\mu} \tilde{Q}_{\mu}^a \ln\left( \frac{\Lambda^2}{\tilde{Q}_{\mu}^b\tilde{Q}_{\mu}^c\bar{\sigma}_b\sigma_c} \right), \\
	&= r^a  + \frac{1}{2}\left[ \sum_i Q^a_i\ln\left(Q_i^b Q_i^c\bar{\sigma}_b\sigma_c\right)  -  \sum_{\mu} \tilde{Q}^a_{\mu} \ln \left(\tilde{Q}_{\mu}^b \tilde{Q}_{\mu}^c \bar{\sigma}_b \sigma_c \right) \right]
\\
& \hspace*{0.75in}
 - \left(\sum_i Q^a_i - \sum_{\mu} \tilde{Q}^a_{\mu} \right) \ln \Lambda,
\end{align*}
where $a = 1, \dots, k$. Introduce the physical scale $\mu$ and from dimension analysis,
\begin{equation*}
	\tilde{Q}_{\mu}^b\tilde{Q}_{\mu}^c\bar{\sigma}_b\sigma_c = C \mu^2,\quad \tilde{Q}_{\mu}^b \tilde{Q}_{\mu}^c \bar{\sigma}_b \sigma_c = \tilde{C} \mu^2,
\end{equation*}
where $C$ and $\tilde{C}$ are nonzero constants. Then from the definition of the beta function, we have 
\begin{equation*}
 \beta^a = \mu \frac{\partial r^a_{\rm eff}}{\partial \mu } = \sum_i Q^a_i - \sum_{\mu} \tilde{Q}^a_{\mu}.
\end{equation*}
This is where we get the constraints Eq.~(\ref{eq:nlsmcd}). 
In particular, if the charges satisfy
\begin{equation}
\label{cd:cy1}
	\sum_i Q^a_i - \sum_{\mu} \tilde{Q}^a_{\mu} = 0,
\end{equation}
$\beta = 0$ and the correction is $\Lambda$ independent, and it gives us a conformal field theory.  When we compare GLSMs for supermanifolds to related GLSMs for hypersurfaces (or complete intersections) in next section, we will see that these conditions correspond to the Calabi-Yau conditions for the hypersurfaces (or complete intersections):
\begin{equation}
\label{cd:cy2}
	\sum_i Q_i^a = \sum_{\mu} \tilde{Q}_{\mu}^a.
\end{equation}
For convenience, we refer to both conditions, 
(\ref{cd:cy1}) and (\ref{cd:cy2}), as the Calabi-Yau condition. 
This is also a hint that indicates there exists a close relationship 
between those two models \cite{Sethi:1994ch,Schwarz:1995ak}.

\subsection{Chiral Ring Relation}
\label{sec:chrialring}
From the effective value of $r$, we could also write down the effective twisted superpotential:
\begin{equation}
\label{eq:correctedtwistsuperpotential}
\tilde{W}_{\rm eff}(\Sigma_a) = - t^a \Sigma_a - \Sigma_a \left[ \sum_i Q_i^a \ln\left( \frac{Q_i^b \Sigma_b}{\Lambda}  \right) - \sum_{\mu} \tilde{Q}_{\mu}^a \ln\left( \frac{\tilde{Q}_{\mu}^b \Sigma_b}{\Lambda}  \right) \right].
\end{equation}

The above one-loop corrected effective twisted potential (\ref{eq:correctedtwistsuperpotential}) can be rewritten in terms of the physical scale \cite{Hori:2003ic}, $\mu$, as
\begin{equation*}
	\tilde{W}_{\rm eff}(\Sigma_a) = - t^a \Sigma_a - \Sigma_a \left[ \sum_i Q^a_i  \left( \ln \frac{Q_i^b \Sigma_b}{\mu} - 1 \right) - \sum_{\nu} \tilde{Q}^a_{\nu} \left(\ln \frac{\tilde{Q}_{\nu}^b\Sigma_b}{\mu}-1\right) \right].
\end{equation*}
The Coulomb branch vacua are found by solving
\begin{equation*}
	\exp\left(\frac{\partial \tilde{W}_{\rm eff}}{\partial \sigma_a} \right)= 1,
\end{equation*}
we can read off the chiral relation as
\begin{equation*}
	q_a \equiv e^{-t_a} = \prod_{i}\left( \frac{Q_i^b \sigma_b}{\mu} \right)^{Q_i^a} \prod_{\nu} \left( \frac{\tilde{Q}_{\nu}^b \sigma_b}{\mu} \right)^{-\tilde{Q}_{\nu}^a}.
\end{equation*}
This is an exact relation where all the $\sigma$'s satisfy. Usually, we set the physical scale $\mu = 1$, then the above relation can be simply written as
\begin{equation}
\label{eq:chiralringrelation}
	q_a  = \prod_{i}\left( Q_i^b \sigma_b \right)^{Q_i^a} \prod_{\nu} \left( \tilde{Q}_{\nu}^b \sigma_b \right)^{-\tilde{Q}_{\nu}^a}.
\end{equation}

We will see in the next section that the GLSM for the hypersurface corresponding to this supermanifold has the chiral ring relation:
\begin{equation}
\label{eq:rechiralringrelation}
	\tilde{q}_a = \prod_{i}\left( Q_i^b \sigma_b \right)^{Q_i^a} \prod_{\nu} \left(-\tilde{Q}_{\nu}^b \sigma_b \right)^{-\tilde{Q}_{\nu}^a},
\end{equation}
It is easy to see that above two chiral ring relations are related by
\begin{equation*}
	q_a = (-1)^{\sum_{\nu} \tilde{Q}_{\nu}^a}\tilde{q}_a.
\end{equation*}
Actually, the factor $(-1)^{\sum_{\nu}\tilde{Q}_{\nu}^a}$ will show up repeatedly in next sections, and we will call this the map connecting the GLSM for a supermainfold to the corresponding GLSM for a hypersurface (or complete intersection).

\subsection{Supersymmetric Localization for Supermanifolds}

In this section, we want to focus on calculations of correlation functions for supermanifolds. Here we only list results of GLSMs for supermanifolds on $S^2$ and it can be generalized to higher genus cases (at fixed complex structure)
as in \cite{Nekrasov:2014xaa,Benini:2016hjo,Closset:2015rna}. Similar to the calculations given in section \ref{sec:glsm}, we could also use supersymmetric localization on Coulomb branches for supermanifolds. However, here we have several Grassmann odd chiral superfields, and they will also contribute to the one-loop determinants of chiral superfields. As we are considering the abelian case in this paper, the one-loop determinants for the gauge fields is trivial by the same argument in \cite{Closset:2015rna,Benini:2015noa}. The one-loop determinant for chiral superfields can be written as the product of even and odd parts: \footnote{This factorization property is still true if we turn on superpotentials, because those higher-order interaction terms appearing in superpotentials will be suppressed by supersymmetric localization. For the same reason, this is also true when discuss about partition functions in section \ref{sec:generalization}.}
\begin{equation*}
	Z_{\mathbf{k}}^{\rm 1-loop} = Z_{\mathbf{k}, {\rm even}}^{\rm 1-loop} \cdot Z_{\mathbf{k}, {\rm odd}}^{\rm 1-loop},
\end{equation*}
where
\begin{subequations}
\label{eq:1loop}
	\begin{align}
	Z_{\mathbf{k}, {\rm even}}^{\rm 1-loop} &= \prod_{i} \left(Q_i^a \sigma_a + \tilde{m}_i \right)^{R_i - 1 - Q_i(\mathbf{k})},  \label{eq:even1loop}\\
	Z_{\mathbf{k}, {\rm odd}}^{\rm 1-loop} &= \prod_{\mu } \left(\tilde{Q}_{\mu}^a \sigma_a + \tilde{m}_{\mu} \right)^{-\tilde{R}_{\mu} + 1 + \tilde{Q}_{\mu}(\mathbf{k})}. \label{eq:odd1loop}
	\end{align}
\end{subequations}

In above, $R_i$ and $\tilde{R}_\mu$ are the $R_V$ charges for 
even chiral superfields and odd chiral superfields, respectively, 
and they are all integers. In Appendix \ref{app:rcharge}, we discuss the assignments of $R_V$-charges. Roughly speaking, except for the P-fields, $R_V$-charges for odd chiral superfields should be proportional to those for even chiral superfields. 
Since we are considering twisted models without superpotentials in this paper, 
specifically without the $P$-fields arising in descriptions of
hypersurfaces, $R_V$-charges for both even and odd chiral superfields should all be assigned to be zero in twisted models. This $R_V$-charge assignment is also consistent with the large volume limit requirement \cite{Herbst:2008jq}.

Before to get the one-loop determinant for odd chiral superfields, (\ref{eq:odd1loop}), let us briefly review the method to obtain (\ref{eq:even1loop}) following \cite{Closset:2015rna,Benini:2015noa}. For Grassmann even superfields $\Phi^i = \left( \tilde{\phi}^i, \psi^i,\dots \right)$, the one-loop determinant from supersymmetric localization is given by
\begin{equation*}
	Z^{\rm 1-loop}_{\rm even} = \prod_i \frac{\det \Delta_{\rm \psi^i}}{\det \Delta_{\rm \phi^i}},
\end{equation*}
where $\det\Delta_{\rm \phi}$ in the denominator comes from the Gaussian integral while $\det \Delta_{\rm \psi}$ in the numerator comes from the Grassmann integral. 
Because of supersymmetry, the only thing that will survive 
from the above ratio is the zero modes of $\psi$, which is (\ref{eq:even1loop}).
It is straightforward to generalize above story for Grassmann odd chiral superfields. For odd chiral superfields $\tilde{\Phi}^\mu = \left( \tilde{\phi}^\mu, \tilde{\psi}^\mu,\dots \right)$, 
the statistical properties of the components $\tilde{\phi}^\mu$ and $\tilde{\psi}^{\mu}$ are exchanged, $\tilde{\phi}^{\mu}$ become Grassmann odd while $\tilde{\psi}^{\mu}$ become Grassmann even. At the same time, the operators, $\Delta_{\tilde{\psi}}$ and $\Delta_{\tilde{\phi}}$, have the same form as those for even chiral superfields \cite{Seki:2005hx}. Therefore, we can use \cite{Closset:2015rna,Benini:2015noa} to get the one-loop determinant for odd chiral superfields:
\begin{equation*}
	Z^{\rm 1-loop}_{\rm odd} = \prod_{\mu} \frac{\det \Delta_{\rm \tilde{\phi}^{\mu}}}{\det \Delta_{\rm \tilde{\psi}^{\mu}}},
\end{equation*}
which leads to (\ref{eq:odd1loop}).

Once we have the one-loop determinant for both even and odd chiral superfields, (\ref{eq:even1loop}) and (\ref{eq:odd1loop}), the correlation function for a general operator $\mathcal{O}(\sigma)$ can also be obtained by
\begin{equation}
\label{eq:crs}
	\left<\mathcal{O}(\sigma)\right> = \sum_{\mathbf{k}}\oint_{\rm JK-Res} \prod_{a=1}^k \left( \frac{{\rm d}\sigma_a}{2\pi i}\right) \mathcal{O}(\sigma) Z_{\mathbf{k}, {\rm even}}^{\rm 1-loop} Z_{\mathbf{k}, {\rm odd}}^{\rm 1-loop} q^{\mathbf{k}},
\end{equation}
Here, the JK-residue calculation is also done at the geometric phase.

\subsection{Elliptic Genera}

The elliptic genus is a powerful tool to extract some physical quantities of a target space, for example the central charge for a Calabi-Yau and the Witten index and so on. It is the partition function on the torus 
with twisted boundary conditions, which reduces to the Witten index in a
certain parameter limit \cite{Witten:1993jg,Gadde:2013ftv,Benini:2013nda}. 
There are many discussions of elliptic genera in the literature.
In this section we will follow the localization computations in \cite{Benini:2013nda,Benini:2013xpa} and generalize their discussions to supermanifolds\footnote{We expect that one can also follow a different approach as in \cite{Gadde:2013ftv} to get a similar result for the supermanifold.}. 
In the next section, we will use our generalizations for supermanifolds to compare to the hypersurface cases, which should provide a 
consistency check that those two models are indeed equivalent to each other under certain conditions. 

In \cite{Benini:2013nda,Benini:2013xpa}, the elliptic genus was computed from supersymmetric localization to be 
\begin{equation*}
	Z_{T^2}(\tau,z) = - \sum_{u_j \in \mathfrak{M}^+_{\rm sing}} \oint_{u=u_j}{\rm d}u \frac{i \eta(q)^3}{\theta_1(q,y^{-1})} \prod_{\Phi_i} \frac{\theta_1(q,y^{R_i/2-1}x^{Q_i})}{\theta_1(q,y^{R_i/2}x^{Q_i})}.
\end{equation*}
Here, we turn off the holonomy of the flavor symmetry on the torus. In the above,
\begin{equation}
\label{eq:xandy}
y = e^{2\pi i z} \quad {\rm and} \quad x_a = e^{2\pi i u_a}
\end{equation}
come from the R symmetry and gauge symmetry, respectively.

The idea is to use supersymmetric localization to transform 
the path integral of a torus partition function 
into a residue integral over zero-modes of vector chiral superfields. In the integrand, the elliptic genus consists of three parts: one-loop determinants for (even) chiral superfields, non-zero modes of vector superfields and twisted chiral superfields. For the supermanifold, we need to include the one-loop determinants for odd chiral superfields with the same twisted boundary conditions on the torus. From supersymmetric localization, the one-loop determinants for odd chiral superfields are almost the same as that for even chiral superfields, except it should have an overall $-1$ exponent.

Now we argue that we would have a very similar formula for elliptic genera for supermanifolds, and the only difference is to include the one-loop determinants for odd chiral superfields. The result is
\begin{equation}
\label{eq:ellipticgenus}
	Z_{T^2}(\tau,z) = - \sum_{u_j \in \mathfrak{M}^+_{\rm sing}} \oint_{u=u_j}{\rm d}u \frac{i \eta(q)^3}{\theta_1(q,y^{-1})} \prod_{\Phi_i} \frac{\theta_1(q,y^{R_i/2-1}x^{Q_i})}{\theta_1(q,y^{R_i/2}x^{Q_i})}\prod_{\tilde{\Phi}_{\mu}}\frac{\theta_1(q,y^{R_{\mu}/2}x^{\tilde{Q}_{\mu}})}{\theta_1(q,y^{R_{\mu}/2-1}x^{\tilde{Q}_{\mu}})}.
\end{equation}

Our argument mainly follows \cite{Benini:2013nda}, and we follow the
notation of that reference. First, we shall note that with twisted boundary conditions on the torus, the one-loop determinants for odd chiral superfields can be calculated from localization:
\begin{equation*}
	Z_{\tilde{\Phi}_{\mu},\tilde{Q}_{\mu}} = \prod_{m,n}\frac{\left|m+n\tau+\frac{R_{\mu}}{2} z + \tilde{Q}_{\mu} u\right|^2 + i \tilde{Q}_{\mu}D}{\left(m+n\tau +(1-\frac{R_{\mu}}{2})z - \tilde{Q}_{\mu}u\right)\left(m+n\bar{\tau} + \frac{R_{\mu}}{2}\bar{z} + \tilde{Q}_{\mu}\bar{u}\right)},
\end{equation*}
and when $D=0$, it can be written in terms of theta functions as inside the 
integral above.

The starting point is
\begin{equation*}
	Z_{T^2} = \int_{\mathbb{R}}{\rm d}D \int_{\mathfrak{M}} {\rm d}^2 u f_{e,g}(u,\bar{u},D)\exp\left[ -\frac{1}{2e^2}D^2 - i\zeta D \right],
\end{equation*}
but with a different $D$-term here, 
which is given in Eq.~(\ref{eq:dterm}). Following the procedure in \cite{Benini:2013nda}, we want to integrate over $D$ and simplify the integral over $u$. 
After introducing odd chiral superfields, we can still take certain 
parameter limits to reduce the integral above to 
$\mathfrak{M} \setminus \Delta_{\epsilon}$ 
and then obtain the residue integral formula. Integrating out $D$, we have
\begin{equation*}
	Z_{T^2} = \int_{\mathfrak{M}}{\rm d}^2u F_{e,g}(u,\bar{u}),
\end{equation*}
with
\begin{align*}
	F_{e,0} =& C_{u,e} \int_{\mathbb{C}^{M_{*}|N_{*}}}{\rm d}^{2M_{*}}\phi_i {\rm d}^{2N_{*}}\tilde{\phi}_{\mu}\exp\left[ -\frac{1}{g}\sum_i|Q_i(u-u_*)|^2|\phi_i|^2 - \frac{1}{g}\sum_{\mu}|\tilde{Q}_{\mu}(u-u_*)|^2|\tilde{\phi}_{\mu}|^2 \right]\\
	 &\times \exp\left[ -\frac{e^2}{2}(  \sum_i Q_i|\phi_i|^2 + \sum_{\mu}\tilde{Q}_{\mu}|\tilde{\phi}_{\mu}|^2 - \zeta)^2 \right].
\end{align*}
Here we use $N_*$ to denote the number of odd chiral superfields which has zero-modes $\tilde{\phi}_{\mu}$ at $u_*$. It is easy to see that the odd chiral superfields do not affect arguments in \cite{Benini:2013nda} as we can expand those odd chiral superfields in the exponent up to linear terms, and the integrals over them are just finite constants before taking the limit $e \rightarrow 0$. Therefore, we shall take $\epsilon \rightarrow 0$ and then $e \rightarrow 0$, also denoted as $\lim_{e,\epsilon \rightarrow 0}$, and then the integral will reduce to
\begin{equation*}
	Z_{T^2} = \lim_{e,\epsilon \rightarrow 0} \int_{\mathfrak{M}\setminus\Delta_{\epsilon}} {\rm d}^2 u F_{e,0}(u,\bar{u}).
\end{equation*}
Once we have the above relation, then following derivations will be the same as in \cite{Benini:2013nda} and we could obtain the formula, 
Eq.~(\ref{eq:ellipticgenus}), for elliptic genera for supermanifolds. 

In principle, we can also turn on the holonomies of the flavor symmetries for GLSMs for supermanifolds on the torus. We will return to this point later. 
Before going to the next section, we shall mention that the elliptic genus 
we calculate here has a natural generalization by including odd chiral 
superfields. The authors are not aware of a corresponding mathematical
notion for supermanifolds, and leave that for future work.

\section{Comparison with GLSMs for Hypersurfaces}
\label{sec:comparison}

The main goal of this section is to reproduce the claim of 
\cite{Sethi:1994ch,Schwarz:1995ak}, namely that an A-twisted NLSM on a supermanifold is equivalent to an A-twisted NLSM on a hypersurface (or a complete intersection). 
Instead of discussing these two NLSMs, we consider the corresponding GLSMs, 
namely GLSMs for supermanifolds and GLSMs for hypersurfaces 
(or complete intersections).
However, here is a subtlety: the GLSM FI parameter $t$ is different from
the NLSM parameter $\tau$, reflecting the difference between algebraic
and flat coordinates. 
They are related by the mirror map \cite{Witten:1993yc,Morrison:1994fr}. 
Therefore, we need to show the mirror map for supermanifolds is the same 
as the mirror map for the corresponding hypersurfaces. 
This is indicated by matching the physical two-sphere partition functions \cite{Jockers:2012dk}. 
We will show this in section~\ref{sec:partitionfunction}.

Before working through concrete calculations, let us argue that our calculations are plausible. As mentioned in section \ref{sec:glsm} and \ref{sec:glsmforsupermanifolds}, GLSMs for supermanifolds we considered in this paper have no superpotentials and so the symmetries for target spaces are all kept, while GLSMs for hypersurfaces will have fewer symmetries. Therefore, there are more twisted mass parameters for the supermanifold case. Further, the statement we want to reproduce is proposed for NLSMs, which correspond to the Higgs branches of GLSMs. However, in this section our calculations are all done on Coulomb branches, for example the correlation functions, (\ref{eq:local}) and (\ref{eq:crs}). Nevertheless, the correlation functions on Higgs branches can be achieved by setting the twisted masses to be zero in the correlation functions on Coulomb branches. Therefore, our results can be used to derive the statement in \cite{Sethi:1994ch,Schwarz:1995ak}.

In last section, when we calculate the one-loop correction, the antisymmetric property for odd chiral superfields leads to a minus sign in front of the correction even though we all assign positive charges for both even and odd chiral superfields at first. This minus sign is essential to the equivalent relations between a GLSM for a supermanifold and for a corresponding hypersurface (or complete intersection).

In the following, we will study some concrete examples. In those examples, it is not necessary to impose the Calabi-Yau conditions (\ref{cd:cy1}). In this sense, we also generalize the statement in \cite{Sethi:1994ch,Schwarz:1995ak} to non-Calabi-Yau cases. What we will use to compare are mainly chiral ring relations, correlation functions and elliptic genera.

\subsection{Hypersurface in \texorpdfstring{$\mathbb{C}\mathbb{P}^N$}{TEXT} vs. \texorpdfstring{$\mathbb{C}\mathbb{P}^{N|1}$}{TEXT}}

First, let us recall the chiral ring relations for the GLSM for the hypersurface case. In this model, we shall introduce the superpotential:
\begin{equation*}
	W = PG(\Phi),
\end{equation*}
where $G(\Phi)$ is a degree $d$ polynomial of $\Phi$'s, and $P$ is a chiral superfield with $U(1)$ charge $-d$ and R-charge $2$. Then the twisted superpotential with one-loop corrections is:
\begin{equation*}
	\tilde{W}= - t \Sigma - \Sigma\left[ (N+1)\left( \ln \frac{\Sigma}{\mu} - 1 \right) - d\left(\ln \frac{-d\Sigma}{\mu}-1\right)\right].
\end{equation*}
From 
\begin{equation*}
	\exp\left(\frac{\partial\tilde{W}}{\partial \sigma}\right) = 1,
\end{equation*}
we obtain
\begin{equation*}
	q \equiv e^{-t} = \left( -\frac{d\sigma}{\mu} \right)^{-d} \left( \frac{\sigma}{\mu} \right)^{N+1} = (-1)^d \left(\frac{d\sigma}{\mu} \right)^{-d}\left( \frac{\sigma}{\mu} \right)^{N+1}.
\end{equation*}
Setting $\mu = 1$, we would get
\begin{equation*}
	q = (-1)^d \left(d\sigma \right)^{-d}\sigma^{N+1}.
\end{equation*}

The corresponding supermanifold model we want to compare to the result above is $\mathbb{C}\mathbb{P}^{N|1}$. We can read the chiral ring relation from Eq. (\ref{eq:chiralringrelation}) with one $U(1)$ and only one odd chiral superfield with $U(1)$ charge $d$,
\begin{equation*}
	q  = \sigma^{N+1} \left( d \sigma \right)^{-d}.
\end{equation*}
Comparing above two chiral ring relations, they are the same up to a factor $(-1)^d$.  

Without loss of generality, we can take $N=4$. We will look at the relation between the correlation functions for GLSMs for hypersurfaces of degree $d$ in $\mathbb{C}\mathbb{P}^4$ and those on $\mathbb{C}\mathbb{P}^{4|1}$, which is defined as in Eq. (\ref{eq:cp41}). In the supermanifold case, we shall have fields with $U(1)$ charges: $(1,1,1,1,1,d)$. Using Eq. (\ref{eq:crs}), we will obtain
\begin{equation}
	\left<\mathcal{O}(\sigma) \right> = \sum_{k \geq 0 } \oint \frac{{\rm d} \sigma}{2\pi i} \frac{\mathcal{O}(\sigma)(d\sigma)^{1 + dk} }{\sigma^{5+5k} }q^k.
\end{equation}
Comparing to the hypersurface case, if we redefine $q$ as
\begin{equation*}
	\tilde{q} = (-1)^d q,
\end{equation*}
then the correlation functions for the supermanifold will be exactly the same as those for the hypersurface.

In particular, if we take $d = 5$, the hypersurface will be the quintic. Correlation functions are
\begin{equation*}
\label{corel:super}
	\left< \mathcal{O}(\sigma) \right> = -\sum_{k \geq 0}\oint \frac{{\rm d} \sigma}{2\pi i} \frac{\mathcal{O}(\sigma) (-5\sigma)^2 }{\sigma^{5+5k}(-5\sigma)^{1-5k}}\tilde{q}^k = - \sum_{k \geq 0}\oint \frac{{\rm d} \sigma}{2\pi i} \frac{\mathcal{O}(\sigma) (-5)^{1+5k} }{\sigma^4}\tilde{q}^k.
\end{equation*}
Then, correspondingly, correlation functions for the supermanifold are:
\begin{equation*}
\label{corel:super2}
	\left< \mathcal{O}(\sigma) \right> = \sum_{k \geq 0} \oint \frac{{\rm d} \sigma}{2\pi i} \frac{\mathcal{O}(\sigma) (5\sigma)^{1+5k} }{\sigma^{5+5k}}q^k = \sum_{k \geq 0} \oint \frac{{\rm d} \sigma}{2\pi i} \frac{\mathcal{O}(\sigma) 5^{1+5k} }{\sigma^4}q^k.
\end{equation*}
We shall see that the $\tilde{q}$ and $q$ are related by
\begin{equation*}
	\tilde{q} = (-1)^{5} q,
\end{equation*}
then it is easy to observe that those correlation functions on both models are exactly the same. It is in this sense that we claim we have reproduced the statement in \cite{Sethi:1994ch,Schwarz:1995ak}. 

Further, we can compare their elliptic genera. The quintic example is already calculated in \cite{Benini:2013nda}, which is 
\begin{equation*}
	Z_{T^2}(\tau,z) = - \frac{i\eta(q)^3}{\theta_1(q,y^{-1})} \oint_{u = 0} {\rm d}u \frac{\theta_1(q,x^{-5})}{\theta_1(q,yx^{-5})} \left(\frac{\theta_1(q,y^{-1}x)}{\theta_1(q,x)} \right)^5.
\end{equation*}
and we can generalize it to a more general hypersurface of degree $d$:
\begin{equation*}
	Z_{T^2}(\tau,z) = - \frac{i\eta(q)^3}{\theta_1(q,y^{-1})} \oint_{u = 0} {\rm d}u \frac{\theta_1(q,x^{-d})}{\theta_1(q,yx^{-d})} \left(\frac{\theta_1(q,y^{-1}x)}{\theta_1(q,x)} \right)^5.
\end{equation*}
For the supermanifold $\mathbb{C}\mathbb{P}^{4|1}$, from the formula (\ref{eq:ellipticgenus}), the elliptic genus is
\begin{equation*}
	Z_{T^2}(\tau,z) = - \frac{i\eta(q)^3}{\theta_1(q,y^{-1})} \oint_{u = 0} {\rm d}u \frac{\theta_1(q,x^{d})}{\theta_1(q,y^{-1}x^{d})} \left(\frac{\theta_1(q,y^{-1}x)}{\theta_1(q,x)} \right)^5.
\end{equation*}
According to the property of the theta function:
\begin{equation}
\label{eq:theta1}
	\theta_1(\tau,x) = - \theta_1(\tau,x^{-1}),
\end{equation}
we conclude that the elliptic genera for both models are exactly the same without turning on the holonomy of the flavor symmetry on the torus. 

As the first example, we have shown the equivalent relations between the GLSM for $\mathbb{C}\mathbb{P}^{N|1}$ and the GLSM for the corresponding hypersurface in $\mathbb{CP}^N$. For elliptic genera, the R charge assignment can be more general which is discussed in appendix \ref{app:rcharge}. We also show in appendix \ref{app:ellipticgenus} that the equivalent relation for their elliptic genera is still valid.

\subsection{Hypersurfaces (Complete Intersections) in \texorpdfstring{$\mathbb{W}\mathbb{P}^N$}{TEXT} vs. \texorpdfstring{$\mathbb{W}\mathbb{P}^{N|1}$}{TEXT} (\texorpdfstring{$\mathbb{W}\mathbb{P}^{N|M}$}{TEXT})}

It turns out that one can repeat the game for more general cases. First, it can be generalized to weighted projective spaces. The supermanifold $\mathbb{W}\mathbb{P}_{[Q_1,\dots,Q_{N+1}|\tilde{Q}]}$ is defined by
\begin{equation*}
	\left\{[ X_1, \dots, X_N, \Theta ]\ |\ (X_1, \dots, X_{N+1}, \Theta )\sim ( \lambda^{Q_1} X_1, \dots, \lambda^{Q_{N+1}} X_{N+1},\lambda^{\tilde{Q}} \Theta )  \right\}.
\end{equation*}
So in the GLSM defined for this supermanifold, we have $N+1$ even directions and $1$ odd direction:
\begin{itemize}
	\item $N+1$ even chiral superfields $X_i$ with $U(1)$ charge $Q_i$ and $R$ charge $0$;
	\item $1$ odd chiral superfield $\Theta$ with $U(1)$ charge $\tilde{Q}$ and $R$ charge $0$;
\end{itemize}
For this GLSM, the chiral ring relation can be read from Eq. (\ref{eq:chiralringrelation}) as
\begin{equation*}
	q  = \prod_{i}\left( Q_i \sigma \right)^{Q_i} \left( \tilde{Q} \sigma\right)^{-\tilde{Q}}.
\end{equation*}

From the previous localization formula, we could obtain the correlation functions:
\begin{equation*}
	\left< \mathcal{O}(\sigma) \right> = \sum_{k \geq 0} \oint \frac{{\rm d} \sigma}{2\pi i} \frac{\mathcal{O}(\sigma) ( \tilde{Q} \sigma)^{1 + \tilde{Q} k} }{\prod_i (Q_i \sigma)^{1+Q_i k} } q^k = \sum_{k \geq 0} \oint \frac{{\rm d} \sigma}{2\pi i} \frac{\mathcal{O}(\sigma) (\tilde{Q} )^{1 + k \tilde{Q} } }{\sigma^{N} \prod_i Q_i^{1 + k Q_i } } q^k.
\end{equation*}

In the above, if we redefine
\begin{equation*}
	\tilde{q} = (-1)^{\tilde{Q}} q,
\end{equation*}
then the above chiral ring relation will be
\begin{equation*}
	\tilde{q}  = \prod_{i}\left( Q_i \sigma \right)^{Q_i} \left( - \tilde{Q} \sigma\right)^{-\tilde{Q}},
\end{equation*}
and the correlation function becomes
\begin{equation*}
	\left< \mathcal{O}(\sigma) \right> = - \sum_{k \geq 0} \oint \frac{{\rm d} \sigma}{2\pi i} \frac{\mathcal{O}(\sigma) ( - \tilde{Q})^{1 + k \tilde{Q} } }{\sigma^{N} \prod_i Q_i^{1 + k Q_i } } {\tilde{q}}^k,
\end{equation*}
which reproduce the chiral ring relation and correlation functions for the GLSM of $U(1)$-gauge on one hypersurface of degree $\tilde{Q}$ in $\mathbb{W}\mathbb{P}_{[Q_1,\dots,Q_{N+1}]}$, which is defined by
\begin{itemize}
	\item $N+1$ even chiral superfields $X_i$ with $U(1)$ charges $Q_i$ and $R$ charges $0$;
	\item $1$ even chiral superfield $P$ with $U(1)$ charge $-\tilde{Q}$ and $R$ charge $2$,
\end{itemize}
together with a superpotential
\begin{equation*}
	W = P G(X_i),
\end{equation*}
where $G(X_i)$ is a holomorphic function in $x_i$' of degree $\tilde{Q}$.

Now, let us compare the elliptic genera for above two GLSMs. For the supermanifold case, it can be calculated using (\ref{eq:ellipticgenus}):
\begin{equation*}
	Z_{T^2} = - \sum_{u_j \in \mathfrak{M}^+_{\rm sing}} \oint_{u=u_j}{\rm d}u \frac{i \eta(q)^3}{\theta_1(q,y^{-1})} \frac{\theta_1(q,x^{\tilde{Q}})}{\theta_1(q,y^{-1}x^{\tilde{Q}}) }\prod_{\Phi_i} \frac{\theta_1(q,y^{-1}x^{Q_i})}{\theta_1(q,x^{Q_i})}.
\end{equation*}
Using the property of $\theta_1$-function (\ref{eq:theta1}), we can rewrite above expression for $Z_{T^2}$ as:
\begin{equation*}
	Z_{T^2} = - \sum_{u_j \in \mathfrak{M}^+_{\rm sing}} \oint_{u=u_j}{\rm d}u \frac{i \eta(q)^3}{\theta_1(q,y^{-1})} \frac{\theta_1(q,x^{-\tilde{Q}})}{\theta_1(q,yx^{-\tilde{Q}}) }\prod_{\Phi_i} \frac{\theta_1(q,y^{-1}x^{Q_i})}{\theta_1(q,x^{Q_i})},
\end{equation*}
which is exactly the elliptic genus for the GLSM for the hypersurface. For Calabi-Yau hypersurfaces, we need to further require the Calabi-Yau conditions (\ref{cd:cy1}) and in this example we have:
\begin{equation*}
	\sum_i Q_i = \tilde{Q}.
\end{equation*}
From above arguments on chiral ring relations, correlation functions and elliptic genera, we shall conclude the statement in \cite{Sethi:1994ch,Schwarz:1995ak} is valid.

Second, there is a similar story when we include more odd chiral superfields. Consider a GLSM for $\mathbb{W}\mathbb{P}^{N|M}_{[Q_1,\dots,Q_{N+1}| \tilde{Q}_1,\dots,\tilde{Q}_M]}$, which is defined by
\begin{equation*}
	\left\{ [X_1,\dots,X_{N+1}, \Theta_1,\dots,\Theta_M] \ |\ (\dots,X_{i},\dots, \Theta_i, \dots) \sim (\dots,\lambda^{Q_i} X_i,\dots,\lambda^{\tilde{Q}_{\mu}} \Theta_{\mu}, \dots) \right\}.
\end{equation*}
For the GLSM for this $\mathbb{W}\mathbb{P}^{N|M}$, matter fields are given as:
\begin{itemize}
	\item $N+1$ even chiral superfields $X_i$ with $U(1)$ charges $Q_i$ and $R$ charges $0$,
	\item $M$ odd chiral superfields $\Theta$ with $U(1)$ charges $\tilde{Q}_{\mu}$ and $R$ charges $0$,
\end{itemize}
and the model we want to compare it to is a GLSM for a complete intersection of $M$ hypersurfaces inside $\mathbb{W}\mathbb{P}^N_{[Q_1,\dots,Q_{N+1}]}$, which is defined by
\begin{itemize}
	\item $N+1$ even chiral superfields $X_i$ with $U(1)$ charges $Q_i$ and $R$ charge $0$,
	\item $M$ even chiral superfields $P_{\mu}$ with $U(1)$ charges $-\tilde{Q}_{\mu}$ and $R$ charges $2$,
\end{itemize}
including the superpotential:
\begin{equation*}
	W = \sum_{\mu} P_{\mu} G_{\mu}(X_i),
\end{equation*}
where $G_{\mu}(X_i)$ is a holomorphic function of degree $\tilde{Q}_{\mu}$.

Starting with the GLSM for $\mathbb{W}\mathbb{P}^{N|M}_{[Q_1,\dots,Q_{N+1}| \tilde{Q}_1,\dots,\tilde{Q}_M]}$, the chiral ring relation is
\begin{equation*}
	q  = \prod_{i}\left( Q_i \sigma \right)^{Q_i} \prod_{\nu} \left( \tilde{Q}_{\nu} \sigma \right)^{-\tilde{Q}_{\nu}}.
\end{equation*}
From the localization formula, we can write the one-loop determinant:
\begin{equation*}
	Z_k^{\rm 1-loop} = \frac{\prod_{\mu} (\tilde{Q}_{\mu}\sigma)^{1+k\tilde{Q}_{\mu}}}{\prod_i (Q_i \sigma)^{1 + k Q_i}} = \frac{1}{\sigma^{N+1 - M+ k(\sum Q_i-\tilde{Q}_{\mu})}}\frac{\prod_{\mu} \tilde{Q}_{\mu}^{1+k\tilde{Q}_{\mu}}}{\prod_i Q_i^{1+k Q_i}},
\end{equation*}
So the correlation function is
\begin{equation*}
	\left< \mathcal{O}(\sigma) \right> = \sum_{k \geq 0} \oint \frac{{\rm d}\sigma}{2\pi i} \frac{\mathcal{O}(\sigma)}{\sigma^{N+1 - M+ k(\sum Q_i-\tilde{Q}_{\mu})}}\frac{\prod_{\mu} \tilde{Q}_{\mu}^{1+k\tilde{Q}_{\mu}}}{\prod_i Q_i^{1+k Q_i}}q^k.
\end{equation*}
If we redefine $q$ inside the residue integral as
\begin{equation}
\label{map}
	\tilde{q} = (-1)^{\sum_{\mu} \tilde{Q}_{\mu}} q,
\end{equation}
we would get the chiral ring relation and the correlation function for 
the GLSM for a complete intersection:
\begin{align*}
	\tilde{q}  &= \prod_{i}\left( Q_i \sigma \right)^{Q_i} \prod_{\nu} \left( - \tilde{Q}_{\nu} \sigma \right)^{-\tilde{Q}_{\nu}},\\
	\left< \mathcal{O}(\sigma) \right> &= (-1)^M \sum_{k \geq 0} \oint \frac{{\rm d}\sigma}{2\pi i} \frac{\mathcal{O}(\sigma)}{\sigma^{N+1 - M+ k(\sum Q_i-\tilde{Q}_{\mu})}}\frac{\prod_{\mu} (-\tilde{Q}_{\mu})^{1+k\tilde{Q}_{\mu}}}{\prod_i Q_i^{1+k Q_i}}q^k.
\end{align*}
Therefore, the equivalent relation we expected still holds here. One more thing we shall mention is that under this redefinition, it will produce an overall factor of $(-1)^M$, which corresponds to the factor of $(-1)^{N_*}$ in Eq. (\ref{eq:local}) \cite{Closset:2015rna}  and it suggests that this shall provide an alternative way to explain the factor $(-1)^{N_*}$ arising in the localization formula in \cite{Closset:2015rna}. So far in the above examples, redefinition of $q$ all have the same form, so it is reasonable to propose that in general a GLSM for a supermanifold $M$ and the corresponding GLSM for a hypersurface (or complete intersection) inside $M_{\rm red}$ are related by Eq. (\ref{map}). 

Also, note that if we require the Calabi-Yau condition:
\begin{equation*}
	\sum_i Q_i = \sum_{\mu} \tilde{Q}_{\mu},
\end{equation*}
then the residue integral over $\sigma$ becomes
\begin{equation*}
	\int {\rm d}\sigma \frac{\mathcal{O}(\sigma)}{\sigma^{N+1 - M}},
\end{equation*}
and so there are only nontrivial correlation functions when
\begin{equation*}
	N \geq M,
\end{equation*}
which makes sense as we would like nontrivial complete intersections of $M$ hypersurfaces insider $\mathbb{W}\mathbb{P}^N$.

Further discussions about elliptic genera for GLSMs for $\mathbb{W}\mathbb{P}^{N|M}$ and for the complete intersection in $\mathbb{W}\mathbb{P}^{N}$ confirms their equivalent relation. The elliptic genus for $\mathbb{W}\mathbb{P}^{N|M}$ is 
\begin{equation*}
	Z_{T^2} = - \sum_{u_j \in \mathfrak{M}^+_{\rm sing}} \oint_{u=u_j}{\rm d}u \frac{i \eta(q)^3}{\theta_1(q,y^{-1})} \prod_{\Phi_i} \frac{\theta_1(q,y^{-1}x^{Q_i})}{\theta_1(q,x^{Q_i})} \prod_{\tilde{\Phi}_{\mu}}\frac{\theta_1(q,x^{\tilde{Q}_{\mu}})}{\theta_1(q,y^{-1}x^{\tilde{Q}_{\mu}}) }.
\end{equation*}
Again, by the property of $\theta_1$-function, the above elliptic genus can also be written as:
\begin{equation*}
	Z_{T^2} = - \sum_{u_j \in \mathfrak{M}^+_{\rm sing}} \oint_{u=u_j}{\rm d}u \frac{i \eta(q)^3}{\theta_1(q,y^{-1})} \prod_{\Phi_i} \frac{\theta_1(q,y^{-1}x^{Q_i})}{\theta_1(q,x^{Q_i})} \prod_{\tilde{\Phi}_{\mu}}\frac{\theta_1(q,x^{-\tilde{Q}_{\mu}})}{\theta_1(q,y x^{-\tilde{Q}_{\mu}}) },
\end{equation*}
which is just the elliptic genus for the GLSM for the complete intersection.

\subsection{Multiple \texorpdfstring{$U(1)$}{TEXT}'s}

Now we want to consider A-twisted GLSMs with multiple $U(1)$ gauge, say $U(1)^k$. Let us look at the GLSM for $X$, which is defined by Eq. (\ref{eq:supertoricvariety}) in section \ref{sec:glsmforsupermanifolds}. Chiral ring relations, correlation functions and the elliptic genera have been already calculated as Eq. (\ref{eq:chiralringrelation}), Eq. (\ref{eq:crs}) and Eq. (\ref{eq:ellipticgenus}), respectively. However, to consider twisted theory we need to set all R-charges assigned to even and odd chiral superfields to be zero, namely,
\begin{equation*}
	R_i = 0,\quad \tilde{R}_{\mu} = 0.
\end{equation*}
Then we have
\begin{align*}
	q_a & = \prod_{i}\left( Q_i^b \sigma_b \right)^{Q_i^a} \prod_{\nu} \left(\tilde{Q}_{\nu}^b \sigma_b \right)^{-\tilde{Q}_{\nu}^a},\\
	 \left< \mathcal{O}(\sigma) \right> & = \sum_{\mathbf{k}}\oint_{\rm JK-Res} \prod_{a=1}^k \left( \frac{{\rm d}\sigma_a}{2\pi i}\right) \mathcal{O}(\sigma) {q_a}^{k_a} \prod_{\Phi_i} \left(Q_i^a \sigma_a \right)^{ - 1 - Q_i^a k_a}   \prod_{\tilde{\Phi}_{\mu} } \left(\tilde{Q}_{\mu}^a \sigma_a \right)^{1 + \tilde{Q}_{\mu}^a k_a}, \\
	 Z_{T^2}(\tau,z) &= - \sum_{u_j \in \mathfrak{M}^+_{\rm sing}} \oint_{u=u_j}{\rm d}u \frac{i \eta(q)^3}{\theta_1(q,y^{-1})} \prod_{\Phi_i} \frac{\theta_1(q,y^{-1}x^{Q_i})}{\theta_1(q,x^{Q_i})}\prod_{\tilde{\Phi}_{\mu}}\frac{\theta_1(q,x^{\tilde{Q}_{\mu}})}{\theta_1(q,y^{-1}x^{\tilde{Q}_{\mu}})}.
\end{align*}

Here we want to compare the model above to the A-twisted GLSM for the complete intersection in $X_{\rm red}$, which is defined by following data:
\begin{itemize}
	\item $N+1$ even superfields $\Phi_i$ with $U(1)^k$ gauge charges $Q_i^a$ and R-charges $0$,
	\item $M$ even superfields $P_{\mu}$ with $U(1)^k$ gauge charges $-\tilde{Q}_\mu^a$ and R-charges $2$.
\end{itemize}
with the superpotential 
\begin{equation*}
	W = \sum_{\mu} P_{\mu} G_{\mu}(\Phi_i),
\end{equation*}
where $G_{\mu}(\Phi_i)$ is a homogeneous polynomial of degree $\tilde{Q}_\mu^a$. Chiral ring relations, correlation functions and elliptic genera can be computed as:
\begin{align*}
	\tilde{q}_a & = \prod_{i}\left( Q_i^b \sigma_b \right)^{Q_i^a} \prod_{\nu} \left(-\tilde{Q}_{\nu}^b \sigma_b \right)^{-\tilde{Q}_{\nu}^a},\\
	 \left< \mathcal{O}(\sigma) \right> & = \sum_{\mathbf{k}}\oint_{\rm JK-Res} \prod_{a=1}^k \left( \frac{{\rm d}\sigma_a}{2\pi i}\right) \mathcal{O}(\sigma)\prod_{i} \left(Q_i^a \sigma_a \right)^{R_i - 1 - Q_i^a k_a} \prod_{\mu } \left(-\tilde{Q}_{\mu}^a \sigma_a \right)^{-\tilde{R}_{\mu} + 1 + \tilde{Q}_{\mu}^a k_a} {\tilde{q}_a}^{k_a}, \\
	 Z_{T^2}(\tau,z) &= - \sum_{u_j \in \mathfrak{M}^+_{\rm sing}} \oint_{u=u_j}{\rm d}u \frac{i \eta(q)^3}{\theta_1(q,y^{-1})} \prod_{\Phi_i} \frac{\theta_1(q,y^{-1}x^{Q_i})}{\theta_1(q,x^{Q_i})} \prod_{P_{\mu}} \frac{\theta_1(q,x^{-\tilde{Q}_{\mu}})}{\theta_1(q,yx^{-\tilde{Q}_{\mu}})}.
\end{align*}
If we redefine
\begin{equation}
\label{eq:redefine}
	\tilde{q}^a = (-1)^{\sum_{\mu} \tilde{Q}_{\mu}^a} q^a,
\end{equation}
then the above two sets of quantities are exactly the same.

\section{Generalizations}
\label{sec:generalization}

So far we have discussed twisted $\mathcal{N} = (2,2)$  abelian GLSMs for supermanifolds without superpotentials. In this section, we want to generalize above discussions.

\subsection{Partition Functions on \texorpdfstring{$S^2$}{Lg}}
\label{sec:partitionfunction}

Beyond chiral ring relations, correlation functions and elliptic genera, 
we also find a similar statement about partition functions. This provides an evidence that the mirror maps for supermanifolds and corresponding hypersurfacs are the same \cite{Jockers:2012dk}.

For GLSMs for ordinary manifolds, already known results show that we could calculate their two-sphere partition functions \cite{Benini:2012ui,Jockers:2012dk}. Here, we focus on the $U(1)$ case and one can easily generalize to the multiple $U(1)$'s. Then the two-sphere partition fucntion is given as in \cite{Benini:2012ui}:
\begin{equation}
\label{eq:partitionfunction}
	Z_{S^2} = \sum_m e^{-i m \theta}\int \frac{{\rm d}\sigma}{2\pi} e^{-4\pi i \xi \sigma} Z_{\Phi}^{\rm 1-loop},
\end{equation}
with the one-loop determinant for (even) chiral superfields:
\begin{equation*}
	Z_{\Phi}^{\rm 1-loop} = \prod_{\Phi_i} \frac{\Gamma\left( \frac{R_i}{2} - i Q_i \sigma - Q_i \frac{m}{2} \right)}{\Gamma\left( 1 - \frac{R_i}{2} + i Q_i \sigma - Q_i \frac{m}{2} \right)},
\end{equation*}
where $Q_i$ and $R_i$ are $U(1)$ gauge charges and R-charges for (even) chiral superfield $\Phi_i$. We would use the R-charge conventions as in appendix \ref{app:rcharge}, i.e. $R_i= \zeta Q_i$. 

Now let us consider the complete intersection. We shall introduce P-fields, say $P_{\mu}$, with $U(1)$ charges $-\tilde{Q}_{\mu}$ and R-charges $2-\zeta \tilde{Q}_{\mu}$, where $\tilde{Q}_{\mu}$ is the degree for corresponding hypersurface. Then the one-loop determinant for $\Phi_i$ and $P_{\mu}$ is
\begin{equation}
\label{eq:1loophypersurface}
	Z_{\Phi,P}^{\rm 1-loop} = \prod_{\Phi_i} \frac{\Gamma\left( Q_i \frac{\zeta }{2} - i Q_i \sigma - Q_i \frac{m}{2} \right)}{\Gamma\left( 1 - Q_i \frac{\zeta }{2} + i Q_i \sigma - Q_i \frac{m}{2} \right)} \prod_{P_{\mu}} \frac{\Gamma\left( 1 - \tilde{Q}_{\mu} \frac{\zeta }{2} + i \tilde{Q}_{\mu} \sigma + \tilde{Q}_{\mu} \frac{m}{2} \right)}{\Gamma\left( \tilde{Q}_{\mu} \frac{\zeta }{2} - i \tilde{Q}_{\mu} \sigma + \tilde{Q}_{\mu} \frac{m}{2} \right)}.
\end{equation}

Here we want to compare it to the partition function for GLSMs for supermanifolds. Therefore, the one-loop determinant for chiral superfields in above partition function should include both even and odd parts. The number of odd chiral superfields, $\tilde{\Phi}_{\mu}$ should be the same as that of P-fields, and $\tilde{\Phi}_{\mu}$ have gauge charge $\tilde{Q}_{\mu}$. From the localization for odd chiral superfields, there should be an overall $-1$ exponent for the one-loop determinant for the odd chiral superfields. Namely, we shall have
\begin{equation*}
	Z_{\Phi}^{\rm 1-loop} = \prod_{\Phi_i} \frac{\Gamma\left( \frac{R_i}{2} - i Q_i \sigma - Q_i \frac{m}{2} \right)}{\Gamma\left( 1 - \frac{R_i}{2} + i Q_i \sigma - Q_i \frac{m}{2} \right)}\prod_{\tilde{\Phi}_{\mu}} \frac{\Gamma\left( 1 - \frac{R_{\mu}}{2} + i \tilde{Q}_{\mu} \sigma - \tilde{Q}_{\mu} \frac{m}{2} \right)}{\Gamma\left( \frac{R_{\mu}}{2} - i \tilde{Q}_{\mu} \sigma - \tilde{Q}_{\mu} \frac{m}{2} \right)}.
\end{equation*}
The partition function would have the form as in Eq.~(\ref{eq:partitionfunction}). Follow the convention in appendix \ref{app:rcharge}, for the supermanifold case, the partition function on $S^2$ is Eq.~(\ref{eq:partitionfunction}) with the one-loop determinant for even and odd chiral superfields:
\begin{equation}
\label{eq:1loopforsuper}
	Z_{\Phi,\tilde{\Phi}}^{\rm 1-loop} = \prod_{\Phi_i} \frac{\Gamma\left( Q_i \frac{\zeta}{2} - i Q_i \sigma - Q_i \frac{m}{2} \right)}{\Gamma\left( 1 - Q_i \frac{\zeta}{2} + i Q_i \sigma - Q_i \frac{m}{2} \right)}\prod_{\tilde{\Phi}_{\mu}} \frac{\Gamma\left( 1 - \tilde{Q}_{\mu}\frac{\zeta}{2} + i \tilde{Q}_{\mu} \sigma - \tilde{Q}_{\mu} \frac{m}{2} \right)}{\Gamma\left( \tilde{Q}_{\mu}\frac{\zeta}{2} - i \tilde{Q}_{\mu} \sigma - \tilde{Q}_{\mu} \frac{m}{2} \right)}.
\end{equation}
From the property of Gamma function:
\begin{equation*}
\Gamma(1-z)\Gamma(z) = \pi/\sin(\pi z)\quad {\rm for}	\quad z \not\in \mathbb{Z}.
\end{equation*}
we know that 
\begin{equation*}
	\frac{\Gamma\left( 1 - \tilde{Q}_{\mu} \frac{\zeta }{2} + i \tilde{Q}_{\mu} \sigma + \tilde{Q}_{\mu} \frac{m}{2} \right)}{\Gamma\left( \tilde{Q}_{\mu} \frac{\zeta }{2} - i \tilde{Q}_{\mu} \sigma + \tilde{Q}_{\mu} \frac{m}{2} \right)} = (-1)^{\tilde{Q}_{\mu} m } \frac{\Gamma\left( 1 - \tilde{Q}_{\mu}\frac{\zeta}{2} + i \tilde{Q}_{\mu} \sigma - \tilde{Q}_{\mu} \frac{m}{2} \right)}{\Gamma\left( \tilde{Q}_{\mu}\frac{\zeta}{2} - i \tilde{Q}_{\mu} \sigma - \tilde{Q}_{\mu} \frac{m}{2} \right)}.
\end{equation*} 
Therefore, we have the following relation between Eq.~(\ref{eq:1loophypersurface}) 
and Eq.~(\ref{eq:1loopforsuper}):
\begin{equation*}
	Z_{\Phi,P}^{\rm 1-loop} = (-1)^{m\sum_{\mu}\tilde{Q}_{\mu}} Z_{\Phi,\tilde{\Phi}}^{\rm 1-loop} .
\end{equation*}
If we shift $\theta$-angle by $\sum_{\mu}\tilde{Q}_{\mu}$ in Eq.(\ref{eq:partitionfunction}), then above factor $(-1)^{m\sum_{\mu}\tilde{Q}_{\mu}}$ can be absorbed inside the sum over $m$, and therefore the partition functions for GLSMs for complete intersections and for corresponding supermanifolds are the same. This shift of $\theta$-angle is nothing but the redefinition of $q$ as we mentioned before in Eq.(\ref{map}). In this sense, it is consistent with discussions in section \ref{sec:comparison}.

\subsection{(0,2) Deformations}
\label{sec:(0,2)}

The calculations in section \ref{sec:comparison} can be extended to $(0,2)$ supersymmetric theories which are deformations of (2,2) theories,
in which case the number of right fermions and left fermions are the same. In particular, we only consider the $E$-deformations here. 
By recent work in $(0,2)$ localization \cite{Closset:2015ohf}, the correlation functions of a general operator $\mathcal{O}(\sigma)$ is given by:
\begin{equation*}
	\left< \mathcal{O}(\sigma) \right> = \sum_{\bm k} \oint_{\rm JKG-Res} \frac{{\rm d}\sigma}{2\pi i} \mathcal{O}(\sigma) Z_{\bm k}^{\rm 1-loop} q^{\bm k}.
\end{equation*}
For the toric case, we have
\begin{equation*}
	Z_{\bm k}^{\rm 1-loop} = \prod_{i} \left(\det M_{i}\right)^{r_{i} - 1 - Q_{i}(\bm k)}.
\end{equation*}
In the above,
\begin{equation*}
	M_{i} = \frac{\partial E_{i}}{\partial \phi},
\end{equation*}
where $E_i$ refer to the $E$-terms as in \cite{Witten:1993yc}.

First，consider a $(2,2)$ GLSM for one hypersurface of degree $(d_1, d_2)$ inside $\mathbb{P}^1 \times \mathbb{P}^1$. The fields and their gauge charges under $U(1) \times U(1)$ are given by
\begin{equation*}
\label{data:3}
\begin{aligned}&
\arraycolsep=3.5pt\def\arraystretch{1.2}
\begin{array}{|cc|cc|c|}
\hline
X_1 &X_2 & Y_1 & Y_2  & P \\ \noalign{\hrule height 0.8pt}\begin{array}{c}
1   \\
0  
\end{array}
&\begin{array}{cc}
1  \\
0  
\end{array}  
&\begin{array}{cc}
0  \\
1
\end{array}&  \begin{array}{cc}
0  \\
1  
\end{array}
&\begin{array}{cc}
- d_1   \\
- d_2
\end{array}  \\
\hline
\end{array}
\end{aligned}\end{equation*}
The R-charge assignment is given by
\begin{equation*}
\label{data:3a}
\begin{aligned}&
\arraycolsep=3.5pt\def\arraystretch{1.2}
\begin{array}{|cc|cc|c|}
\hline
X_1 &X_2 & Y_1 & Y_2  & P \\ \noalign{\hrule height 0.8pt}\begin{array}{c}
0  
\end{array}
&\begin{array}{cc}
0  
\end{array}  
&\begin{array}{cc}
0 
\end{array}&  \begin{array}{cc}
0   
\end{array}
&\begin{array}{cc}
2
\end{array}  \\
\hline
\end{array}
\end{aligned}\end{equation*}
The superpotential is 
\begin{equation}
	W = P G(X,Y),
\end{equation}
where $G(X,Y)$ is a homogeneous polynomial of degree $d_1$ in $X_i$ and degree $d_2$ in $Y_i$.

For this case, if written in $(0,2)$ language, the $E_{i}$ are given by
\begin{equation*}
	E_{X_i} = \sigma_1 X_i, \quad
	E_{Y_i} = \sigma_2 Y_i, \quad
	E_P = - d_1 \sigma_1 P - d_2 \sigma_2 P.
\end{equation*}
Therefore,
\begin{equation*}
	M_1 = \sigma_1 \mathbbm{1}_{2\times 2},\quad M_2 = \sigma_2 \mathbbm{1}_{2\times 2}, \quad M_P =  - d_1 \sigma_1 - d_2 \sigma_2.
\end{equation*}
From the $(0,2)$ superpotential, the $J$-terms are
\begin{equation*}
	J_{X_i} = P \frac{\partial G}{\partial X_i},\quad J_{Y_i} = P\frac{\partial G}{\partial Y_i},\quad J_P = G(X,Y).
\end{equation*}

Now, consider $(0,2)$ deformations of the above model.
For simplicity we keep all $J$-terms undeformed and $E_P$ undeformed. 
In general, the $E$-deformations written in matrix form are
\begin{equation}
\label{eq:deformation}
	E_X = \sigma_1 A X + \sigma_2 B X,\quad E_Y = \sigma_1 C Y + \sigma_2 D Y.
\end{equation}
(See e.g. \cite{McOrist:2008ji,Donagi:2011uz,Donagi:2011va} for a discussion of $(0,2)$ deformations of tangent bundles
of products of projective spaces and results in quantum sheaf cohomology.)
Then the $M$'s are given by:
\begin{equation}
	M_X = A \sigma_1 + B \sigma_2,\quad M_Y = C \sigma_1 + D \sigma_2.	
\end{equation}
In the above, $A$, $B$, $C$ and $D$ are $2 \times 2$ matrices. 
For simplicity, we shall require $A$ and $D$ are invertible, while $B$ and $C$ are not. Furthermore, supersymmetry requires $E\cdot J = 0$, 
therefore the matrices above satisfy following constraints \cite{Closset:2015ohf,Parsian:2018fhm}:
\begin{subequations}
	\begin{align}
		\frac{\partial G}{\partial X_i} \left( A_{ij} - \delta_{ij}\right) X_j + \frac{\partial G}{\partial Y_i} C_{ij}Y_j &=0,   \label{eq:susyconstraint1} \\
		\frac{\partial G}{\partial X_i} B_{ij} X_j + \frac{\partial G}{\partial Y_i} \left( D_{ij} - \delta_{ij}\right)Y_j &=0.   \label{eq:susyconstraint2} 
	\end{align}
\end{subequations}
It is easy to see that there is a special solution to the equations above: 
take $A$ and $D$ to be the identity and $B$ and $C$ to be zero.
This corresponds to the $(2,2)$ case.

From the localization formula in \cite{Closset:2015ohf}, we have
\begin{equation}
\label{eq:02correlationfunction}
	\mathcal{O}(\sigma_1,\sigma_2) = (-1) \sum_{k_1,k_2}\oint_{JKG-Res} \frac{{\rm d}\sigma_1}{2\pi i} \wedge \frac{{\rm d}\sigma_2}{2\pi i}\mathcal{O}(\sigma_1,\sigma_2)\frac{(-d_1 \sigma_1 - d_2 \sigma_2)^{1+d_1k_1+d_2k_2}}{(\det M_X)^{1+k_1}(\det M_Y)^{1+k_2}} \tilde{q}_1^{k_1} \tilde{q}_2^{k_2}.
\end{equation}

From section \ref{sec:glsmforsupermanifolds} and \ref{sec:comparison}, there is a corresponding story in the supermanifold case. The GLSM for the corresponding supermanifold is given by following data:
\begin{equation*}
\label{data:3b}
\begin{aligned}&
\arraycolsep=3.5pt\def\arraystretch{1.2}
\begin{array}{|cc|cc|c|}
\hline
X_1 &X_2 & Y_1 & Y_2  & \theta \\ \noalign{\hrule height 0.6pt}\begin{array}{c}
1   \\
0  
\end{array}
&\begin{array}{cc}
1  \\
0  
\end{array}  
&\begin{array}{cc}
0  \\
1
\end{array}&  \begin{array}{cc}
0  \\
1  
\end{array}
&\begin{array}{cc}
 d_1   \\
 d_2
\end{array}  \\
\hline
\end{array}
\end{aligned}\end{equation*}
with all R charges vanishing, and there is no superpotential. 
As a result, $J = 0$ and so $E\cdot J = 0$ trivially.  
Therefore, in the supermanifold case, there is no constriant on $A, B, C, D$.

We also keep the $E_{\theta}$ term undeformed for simplicity:
\begin{equation*}
	E_{\theta} = d_1 \sigma_1 \theta + d_2 \sigma_2 \theta, \quad M_{\theta}= d_1 \sigma_1 + d_2 \sigma_2.
\end{equation*}
Following the same argument in section \ref{sec:glsm}, the general correlation function is given as
\begin{equation}
\label{eq:02correlationfunctionsuper}
	\mathcal{O}(\sigma_1,\sigma_2) = \sum_{k_1,k_2}\oint_{JKG-Res} \frac{{\rm d}\sigma_1}{2\pi i} \wedge \frac{{\rm d}\sigma_2}{2\pi i}\mathcal{O}(\sigma_1,\sigma_2)\frac{(d_1 \sigma_1 + d_2 \sigma_2)^{1+d_1k_1+d_2k_2}}{(\det M_X)^{1+k_1}(\det M_Y)^{1+k_2}} q_1^{k_1} q_2^{k_2}.
\end{equation}
The expression of (\ref{eq:02correlationfunction}) and (\ref{eq:02correlationfunctionsuper}) are related by Eq. (\ref{eq:redefine}). Those correlation functions are exactly the same only when (\ref{eq:susyconstraint1}) and (\ref{eq:susyconstraint2}) are satisfied. However, we should emphasize that GLSMs for supermanifolds admits more $(0,2)$ deformations.

In this section we have only considered a simple example and it can be generalized to more general cases. Therefore, we would like to conjecture that there exists an $(0,2)$ analogue of the statement about supermanifolds in \cite{Sethi:1994ch,Schwarz:1995ak}: under certain constraints on $(0,2)$ deformation, an A/2-twisted NLSM on a hypersurface or complete intersection \cite{Katz:2004nn,Melnikov:2012hk} is equivalent to an A/2-twisted NLSM on some supermanifold.

\section{Conclusions}

In this paper we have found evidences in GLSMs for
the relation described in \cite{Sethi:1994ch,Schwarz:1995ak} between
sigma models on supermanifolds and hypersurfaces, by using the supersymmetric 
localization. We also find a similar relationship for elliptic genera
of supermanifolds and hypersurfaces, and also in $(0,2)$ deformations of
supermanifolds and hypersurfaces.

Another possible future direction is to understand
mirror symmetry for supermanifolds. 
Some previous studies exist \cite{Aganagic:2004yh,Seki:2005hx}, and it may be possible to make further progress using
supersymmetric localization as in \cite{Gu:2017nye,Gu:2018fpm}.

\section*{Acknowledgment}

We would like to thank Cyril Closset and Jirui Guo for reading the manuscript and useful comments.  We thank in particular Eric Sharpe for collaborations at the beginning of this project, many useful discussions and helpful suggestions regarding the writing.

\appendix
\addcontentsline{toc}{section}{Appendices}

\section{Vector R-charges}
\label{app:rcharge}

In this section, we will discuss the assignment of R-charges to chiral superfields in physical models, especially for odd chiral superfields.

For A-twisted models without superpotentials (e.g. without $P$-fields), we always assign vanishing R-charges to chiral superfields $\Phi_i$'s. If the superpotentials are nonzero, then they must have total R-charges two, so one must assign nonzero R-charges to some of the chiral superfields.

First consider all chiral superfield $\Phi_i$ are charged under only one $U(1)$ gauge symmetry.  We can mix $U(1)_R$ with this $U(1)$ to get a 
new $U(1)^{\prime}_{R}$ R-symmetry \cite{Benini:2013nda,Benini:2016qnm}:
\begin{equation*}
	U(1)^{\prime}_R = U(1)_R + \zeta U(1),
\end{equation*}
where $\zeta$ is the deformation parameter. After mixing, the new $U(1)$ R-charge is
\begin{equation*}
	R_i^{\prime} = R_i + \zeta Q_i.
\end{equation*}
If starting with $R_i=0$, we can continuously deform it to be $R_i^{\prime} = \zeta Q_i$ as the new R-charge. Therefore, nonzero R-charges assigned to (even) chiral superfields should be proportional to their weights. For convenience, we will denote $R_i^{\prime}$ also as $R_i$ following without causing any confusion. Thus, the R-charges are assigned to be:
\begin{equation*}
	R_i = \zeta Q_i.
\end{equation*}
Now consider the $P$ field, in the superpotential $W = PG(\Phi)$, 
where $G(\Phi)$ is a degree $d$ polynomial in $\Phi_i$'s.  
\begin{equation*}
	d = \sum_i n_i Q_i,
\end{equation*}
for a set of integers $\{n_i\}$ and $n_i$ comes from the power of $\Phi_i$ 
in one term of the (quasi-)homogeneous polynomial $G$. 
Then the $U(1)$ charge for this $P$-field should be $-d$. 
To guarantee $R_{W} = 2$, we need to assign the $P$ field R-charge:
\begin{equation*}
	R_{P} = 2 - \sum_{i}n_i R_i =2 - \zeta\sum_{i} n_i Q_i  = 2 - \zeta d.
\end{equation*}
In the above, 
when $\zeta = 0$, it agrees with the assignments in A-twisted models.

In the toric supermanifold case, 
odd chiral superfields and even chiral superfields share the same $U(1)$ gauge, and so we should assign R charges to those odd chiral superfields by:
\begin{equation*}
	\tilde{R}_{\mu} = \zeta \tilde{Q}_{\mu}.
\end{equation*}
Specifically, if we consider A-twisted theories, R charges should be assigned as
\begin{equation*}
	R_i = 0, \quad {\rm and}\quad \tilde{R}_{\mu} = 0. 
\end{equation*}

These computations can be generalized to multiple $U(1)$'s.

\section{Lagrangian on Curved Spaces}

\label{app:curvedspacecorrection}

In section \ref{sec:themodel}, we described GLSMs for supermanifolds on 
flat worldsheets. However, in this paper we also consider GLSMs 
for supermanifolds on the two-sphere. 
Since $S^2$ is not flat, the Lagrangian will have curvature correction 
terms \cite{Benini:2012ui,Doroud:2012xw,Closset:2014pda}. 
In this section, we want to write out Lagrangians for GLSMs 
for supermanifolds on a worldsheet two-sphere. 
Since the only difference with GLSMs for ordinary spaces is the kinetic term for odd chiral superfields (\ref{eq:oddkin}), we will only write out $\mathcal{L}_{\rm kin}^{\rm odd}$. 

First, consider the physical Lagrangian on $S^2$. By solving the supergravity background, one can follow \cite{Benini:2012ui} to get the kinetic term for the odd superfield $\tilde{\Phi}$ with vector R-charge $\tilde{R}$ as:\footnote{There is another supergravity background used in \cite{Doroud:2012xw}. These two supergravity backgrounds are claimed to be equivalent to each other as studied in \cite{Closset:2014pda} }
\begin{multline}
	\mathcal{L}_{\rm kin}^{\rm odd} = D_{\mu} \bar{\tilde{\phi}} D^{\mu} \tilde{\phi} + \bar{\tilde{\phi}} \sigma^2 \tilde{\phi} + \bar{\tilde{\phi}} \eta^2 \tilde{\phi} + i \bar{\tilde{\phi}} \tilde{D} \tilde{\phi}  + \bar{\tilde{F}}\tilde{F} + \frac{i \tilde{R}}{r} \bar{\tilde{\phi}} \sigma \tilde{\phi} + \frac{\tilde{R}(2-\tilde{R})}{4r^2} \bar{\tilde{\phi}}\tilde{\phi} \\
	- i\bar{\tilde{\psi}} \gamma^{\mu} D_{\mu} \tilde{\psi} +i \bar{\tilde{\psi}} \sigma \tilde{\psi} - \bar{\tilde{\psi}} \gamma_3 \eta \tilde{\psi} + i \bar{\tilde{\psi}}\tilde{\lambda}\tilde{\phi} - i \bar{\tilde{\phi}}\bar{\tilde{\lambda}}\tilde{\psi} - \frac{\tilde{R}}{2r} \bar{\tilde{\psi}}\tilde{\psi}.
\end{multline}

Similarly, we can follow \cite{Closset:2015rna} to get the twisted Lagrangian on $S^2$. The kinetic term for odd chiral superfields will have the same form as Eq. (2.35) in \cite{Closset:2015rna}. One difference is that the statistical properties for each component field are changed.

\section{Elliptic Genera with General R Charges}
\label{app:ellipticgenus}

In this section, we calculate the elliptic genera for more general R-charge assignments, following Appendix \ref{app:rcharge}. In the same spirit of Section \ref{sec:comparison}, we focus on comparison of hypersurfaces and supermanifolds.

As an example, we only consider the GLSM for the hypersurface in $\mathbb{W}\mathbb{P}^N_{[Q_1,\ldots,Q_{M+1}]}$ and for $\mathbb{W}\mathbb{P}^{N+1|M}_{[Q_1,\ldots,Q_{M+1}|\tilde{Q}]}$. Actually, we only need compare the one-loop determinants for $P$-field, say $P$ with $U(1)$ charge $-\tilde{Q}$, and that for the odd chiral superfield, say $\Psi$ with $U(1)$ charge $\tilde{Q}$. From appendix \ref{app:rcharge}, the R-charge for $P$ is $2-\zeta \tilde{Q}$ and the R-charge for $\Psi$ is $\zeta \tilde{Q}$. Then we have
\begin{align*}
	Z_{P}^{\rm 1-loop} &= \frac{\theta_1(q,y^{R_P/2-1}x^{-\tilde{Q}})}{\theta_1(q,y^{R_P/2}x^{-\tilde{Q}})} = \frac{\theta_1(q,y^{-\zeta \tilde{Q}/2}x^{-\tilde{Q}})}{\theta_1(q,y^{1-\zeta \tilde{Q}/2}x^{-\tilde{Q}})}, \\
	Z_{\Psi}^{\rm 1-loop} &= \frac{\theta_1(q,y^{R_{\Psi}/2}x^{-\tilde{Q}})}{\theta_1(q,y^{R_{\Psi}/2-1}x^{-\tilde{Q}})} = \frac{\theta_1(q,y^{\zeta \tilde{Q}}x^{-\tilde{Q}})}{\theta_1(q,y^{\zeta \tilde{Q}/2 - 1}x^{-\tilde{Q}})}.
\end{align*}
Then according to the property of $\theta_1$-function, $\theta_1(\tau,x) = - \theta_1(\tau,x^{-1})$, above two one-loop determinants equal to each and so do their elliptic genera. This calculation can be easily generalized to more general cases as in section \ref{sec:comparison}.

\phantomsection

\end{document}